# Enhancing Critical Infrastructure Cybersecurity: Collaborative DNN Synthesis in the Cloud Continuum[*]

Lav Gupta
Guoxing Yao

Department of Computer Science, University of Missouri St. Louis

ABSTRACT

Researchers are exploring the integration of IoT and the cloud continuum, together with AI to enhance the cost-effectiveness and efficiency of critical infrastructure (CI) systems cost. This integration, however, increases susceptibility of CI systems to cyberattacks, potentially leading to disruptions like power outages, oil spills, or even a nuclear mishap. CI systems are inherently complex and generate vast amounts of heterogeneous and high-dimensional data, which crosses many trust boundaries in their journey across the IoT, edge, and cloud domains over the communication network interconnecting them. As a result, they face expanded attack surfaces. To ensure the security of these dataflows, researchers have used deep neural network models with encouraging results. Nevertheless, two important challenges that remain are tackling the computational complexity of these models to reduce convergence times and preserving the accuracy of detection of integrity-violating intrusions. In this paper, we propose an innovative approach that utilizes trained edge cloud models to synthesize central cloud models, effectively overcoming these challenges. We empirically validate the effectiveness of the proposed method by comparing it with traditional centralized and distributed techniques, including a contemporary collaborative technique.

1. Introduction

Critical infrastructure (CI) systems form a complex, interconnected ecosystem that is essential for maintaining normalcy in daily life. The Cybersecurity and Infrastructure Security Agency (CISA), the operational part of the Department of Homeland Security of the USA, has identified sixteen key CI sectors encompassing vital domains, including healthcare, communications, financial systems, and transportation. Any threat to the network and other assets, or their management control systems (MCS), endangers national interests [1]. The complexity of safeguarding critical infrastructure has been highlighted by a World Bank report, which states that despite substantial attention and financial investments, protecting critical infrastructure remains an ongoing challenge [2]. Cases in point are the Amsterdam-Rotterdam-Antwerp and the Colonial Pipeline attacks. These high-profile incidents emphasize the need for continuous improvement and innovation in the security of CI systems.

Recent studies reveal an increasing use of the Internet of Things (IoT) and other evolving technologies like edge-computing and AI in the management of critical infrastructure systems [3], [4]. IoT devices form an important part of operational technology (OT) that interfaces with the physical environment of critical infrastructure systems. As part of OT, IoT generates a large amount of data that is crucial for improving the maintenance and scalability of infrastructure assets. IoT helps integrate the physical world of OT with the digital world of information technology (IT), which is proving to be a powerful and cost-effective way to automate operations, management, and control of CI. IT provides flexible processing and storage at the edge and in the central clouds, which makes up for IoT devices being deficient in these resources. Within the realm of IT, central clouds offer flexible resource allocation, lower costs, and protection against resource malfunction. However, the issue of high latency can arise because of factors like remoteness, network congestion, and cloud outages [5]. On the other hand, edge servers and edge clouds, being proximal to the data generation points, provide low turnaround times.

The synergistic use of AI, along with an edge-cloud approach (the term edge-cloud refers to the collaborative deployment across edge and central clouds referred to as the cloud continuum), gives the ability to handle extensive data volumes and provides predictive capabilities necessary for proactive management and control of CI. However, given the current state of development, it would be incorrect to assume that complex AI models can be trained and used at the same level of sophistication and with the same ease at the IoT devices or at the edge as they can be at the central clouds. Researchers and industry professionals now agree that collaborative approaches with deep neural network (DNN) models running at different levels, specifically on both edge and central clouds, can provide useful solutions. The collaborative effort aims to enhance the intelligence of CI systems [6], [7]. Despite the appeal of AI in critical sectors like healthcare and transportation, substantial research challenges persist. The increasing size and complexity of DNN model as we move across the cloud continuum from the IoT domain through the smaller edge clouds to the larger core clouds, result in prolonged training convergence times for deeper central cloud models. This impediment makes it difficult to deploy DNN models for the management of critical real-time and near-real-time services. Efforts to expedite training in real-world systems often lead to a reduction in prediction accuracy. As we navigate these challenges, the collaborative approach remains a promising method for enhancing the security and resilience of CI systems.

---

[*] 

A known issue with the use of deep learning models across the cloud continuum to handle CI workloads is the significant amplification of the attack surface, rendering them susceptible to more severe attacks than traditional systems [8]. These attacks can infiltrate the system, damage the integrity of data streams passing through the IoT-edge-cloud hierarchy, and lead to CI malfunctions like power outages, breakdowns in healthcare systems, oil spills, damage to nuclear centrifuges, and shutdowns of industrial furnaces [9], [10]. To address the security concerns of CI workflows, this paper proposes an innovative approach for collaborative construction and training of deep neural network models working across the edge and the central clouds. Our findings show that the challenges of convergence and accuracy are effectively addressed by the approach that we propose in this paper. The specific contributions of this paper are as follows:

1. *Abstracting a conceptual model for managing IoT-enabled critical infrastructure systems using standard, established, and available frameworks and current research. This model serves as the foundational basis for establishing an AI-powered architectural framework for the security of critical infrastructure systems deployed over the edge-cloud continuum.*
2. *The architecture described above leads to one of the key contributions of this paper. We introduce a novel technique for the deployment of a distributed edge-cloud system of deep neural network models, built and developed in a bottom-up manner. This innovative approach involves the deployment of collaboratively synthesized system edge-cloud deep neural networks that effectively address the challenges of training convergence times and the security of interdomain CI dataflows.*
3. *Providing theoretical and mathematical discussion on the contributions of various layers to the training performance of neural network models. This discussion forms the basis for the selection of layers for synthesizing central cloud models from trained edge-cloud models. We also discuss the computational complexity of DNN models to appreciate how the proposed method achieves its objectives.*
4. *Empirical validation of constituent individual aspects that have great bearing on our proposed synthesis method. These include layer-wise analysis and selection and the ability of the proposed security system to achieve acceleration in training convergence while preserving high accuracy in detecting dataflow integrity violations.*

Through these contributions, our paper seeks to provide a comprehensive and effective approach to enhance the security and resilience of CI systems, effectively tackling the challenges that arise with the use of IoT, edge, and central clouds. The rest of the paper is organized as follows: Section 2 examines key related work and brings out the existing research gaps; Section 3 brings out the challenges of using deep learning models for the protection of CI systems and the motivation for this work; Section 4 discusses the design of the system; and Section 5 empirically validates the proposed methods. Section 6 discusses the result and draws conclusions.

2. Related Works: Current Status and Research Gaps

In this section, we discuss several works that involve the processing of data generated by sensors and other IoT devices in the edge and central clouds. We use this discussion to bring out major gaps in the IoT-edge-cloud-AI research and the significance of our work in addressing some of these gaps.

*2.1 IoT-Edge-AI*

In resource constrained IoT environments where flexible allocation of processing resources from edge servers or edge clouds is necessary, the distribution of data across trusted IoT devices and edge servers is a favored option. Addressing this, the authors in [11] advocate for sensor devices to offload data to edge servers or edge clouds, leveraging the deep reinforcement method for parallel data processing to effectively reduce latency. The researchers in [12] investigate the current requirements of resources when developing edge AI applications. In their discussion, they include the edge-AI applications, the software used to develop AI models, and the hardware products used. The authors in [13] also reinforce ultra-low latency and add saving energy for end devices as prominent advantages of edge computing. In terms of verticals that highlight the usefulness of IoT and edge processing, these include such applications as smart healthcare [11], farming [14], and medical diagnostics [15].

More recently, researchers have stressed the importance of distributed processing in the IoT-edge environment. An example of such an environment is smart factories, in which a wide variety of edge devices are deployed to carry out tasks collaboratively in a heterogeneous and distributed manner. In [16], the researchers propose a collaborative system that processes tasks jointly in a decentralized manner and permits real-time data analysis. Since the training tasks in such applications can be overwhelmed by massive computations and frequent communication, they suggest a system that discovers and uses the available resources of edge devices. The authors define clusters of edge devices that handle one deep learning task independently. Workload is distributed across communication intensive workers (WK) and computation intensive parametric servers (PS) instantiated on various edge devices. WKs retrieve the latest instance of the model from PS, train them on local data subset $\mathbf{D^i}$ by executing local computation $\mathbf{f(.)}$ and obtain partial gradients at time $\mathbf{(t)}$ as $\mathbf{g_i^t = f(w^{t-1}, D^i)}$, where $\mathbf{w^{t-1}}$ is the model state at time $\mathbf{(t-1)}$. These partial gradients are aggregated by the PS with the back propagation function $\mathbf{F(.)}$ to update the model. The convergence criteria used is $\mathbf{w^t = F(w^{t-1}, \sum_i g_i^t)}$. Unprocessed data is jointly handled by WKs according to their computing capability and network conditions.

A standardized implementation of edge, especially in conjunction with 5G cellular networks is that of MEC (multi-access edge computing). The MEC reference architecture is defined in ETSI GS MEC 003 [i.5]. MEC consists of functions at the host and system levels. Host-level functions include the MEC Platform, MEC apps, and Virtualization Infrastructure. Host level management functions include MEC Platform Manager and Virtualization Infrastructure Manager. System level functions include the MEC Orchestrator and the OSS function. When MEC is integrated into the 5G cellular system, the key definitions of MEC in ETSI GS MEC 003 [i.5] should be maintained.

MEC facilitates edge computing with autonomous servers deployed at base stations in diverse locations. It enables the implementation of delay-sensitive and context-aware services. Mobile Edge Computing (MEC) has emerged as a viable solution for reducing the workload being sent to the core network by deploying computing and storage resources at the edge of the network. MEC plays an essential role in the 5G infrastructure since many 5G usage scenarios, viz., augmented reality (AR), video analytics, and content distribution, require edge computing. MEC is expected be natively integrated in 6G cellular and beyond. In [17], the authors have described their IoT-edge implementation in the mobile edge cloud computing environment. In their opinion, system efficiency is enhanced by applying the caching and processing abilities of MEC. As far as the AI models are concerned, they have used restricted Boltzmann machines (RBM) and Deep belief networks optimized by Barnacle Mating Optimizer [17]. The authors in [18] are of the opinion that computational offloading is one of the key technologies of MEC. The MCS should be able to decide which tasks will be handled on the local device or MEC server. The MCS also makes decisions about the processing, storage, and communication resources required to process tasks. Fig. 1 illustrates an ETSI reference deployment scenario for the use of MEC in industrial automation [19].

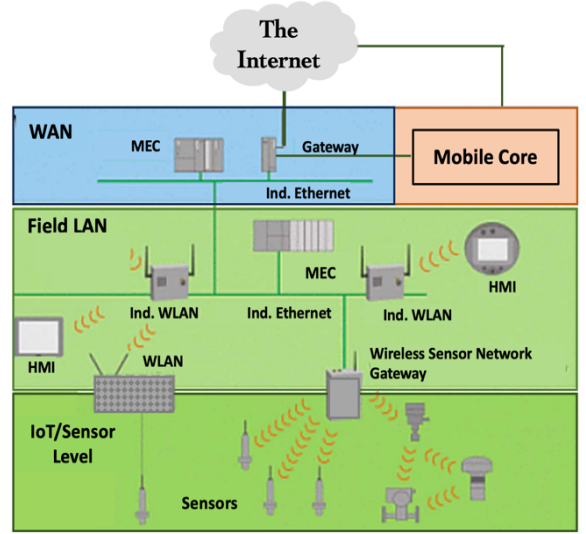

**Fig. 1**. Multi-access edge in industrial automation (adapted from ETSI GS MEC 002 V3.2.1 (2024-02)

However, it is widely believed that there are prominent challenges in edge computing. In [12], authors find Edge AI to be a still-developing ecosystem with its own challenges, such as latency and limited communication resources. Another important issue is maintaining privacy when uploading data to cloud-based servers. In their opinion, deploying complete AI models on edge devices is generally impracticable. In [13], some of the open challenges mentioned by the authors are data scarcity and lack of data consistency because of sensor heterogeneity, which result in low performance. If models are trained on a central server and deployed on edge devices, the trained models do not generalize to unknown data, which results in low performance.

*2.2 Collaborative IoT-Edge-Central Cloud AI*

The IoT-edge computing described in the previous section solves several problems. Some of the important ones are privacy and security issues, providing low latency processing for real-time or near real-time services and avoiding the bandwidth limitations and congestion in wide area communication networks. However, there remains a significant limitation: they do not have sufficient resources to optimize, store, update, and transmit large neural network models. In this context, a computing paradigm shift that is underway is centered around the seamless integration of IoT devices, edges, and central cloud resources, enabled by modern communication networks, in one computing system, referred to as the IoT-edge-cloud continuum or simply the continuum. This the more generalized form of the continuum that many works refer to as the edge-cloud continuum. This continuum is pushing the limits of traditional centralized cloud computing solutions, enabling, among other features, efficient data processing and storage as well as low latency for service execution [20]. The dynamic distribution of workloads over the continuum is expected to provide the much-needed low latency, computation intensive processing required for critical applications [21]. The researchers in this work, however, stress that for this to happen, innovative techniques are required for task offloading to optimize service execution, security and privacy of data, and optimal resource allocation mechanisms.

The researchers in [22] classify the resources available on the computing continuum into computing, networking, and storage. They propose a UNITE framework that evaluates KPIs like latency and throughput. If any of these KPIs fall below a predefined value, appropriate actions will take place. Being constrained in their capabilities, the edge devices are suited to handle simple machine inference tasks, while more complex neural network models need to be trained by the edge or the central cloud resources.

Some issues have been reinforced and new issues have been discussed by the authors in [23]. Besides slower processing times, edge computing has higher cost in terms of energy consumption compared to central servers. Taking cognizance of limited resources at the edge, some authors recommend edge processing only in cases for which the edge devices have sufficient computational capacity [24], [25]. Another practical approach suggested is to execute simpler cases on the edge and offload the difficult ones to the central cloud [26]. Researchers have also suggested the strategy of compressing and executing models at the edge if network conditions are unfavorable and taking them to the central cloud otherwise [13], [23].

To avoid issues associated with centralized training, an alternative approach involves creating a model at the central cloud that learns from the models at the edge nodes [27] [28]. Guo (2022) et al. address federated learning (FL) in their work, wherein a baseline model is created at the central cloud and shared with the edge nodes. The edge nodes train the models with the locally generated data. The edge nodes share the parameters of their models with the central cloud model. The central cloud model uses these parameters to generate new intelligence. The central cloud then shares the updated model with the edge clouds for the next iteration [29].

In their work on the challenges and opportunities of federated learning [30], Ding et al., have highlighted two critical challenges of FL. In systems involving continuous execution, the local training at client nodes involves continuously observed new data as well as the stored static dataset. This increases the convergence time of an FL model. The authors use a binary classification task to illustrate the processes of prediction and detection in these models. Let the training data $(Y_i, X_i), i = 1, \ldots, n$, and test data $(Y, X)$ be IID random variables with values in $\mathbf{R}^d \times \{1, -1\}$. For detection, the goal is to test **H0: y = −1 against H1: y = 1, given x.** Given a fixed false-detection rate **P(f(x) > 0 | y = −1),** we want to minimize the false-rejection rate **P(f(x) < 0 | y = 1).** The authors conclude that FL that averages over many local updates can still converge to a desirable model, while aggressive local updates are likely to harm performance. The limits of communication rounds needed for rate-optimality in FL (if it exists) have yet to be studied. Conventional FL systems are designed to learn only a prespecified task using data collected from a fixed channel. Thus, their generalization capability is quite limited for most real-world learning situations where the underlying data and tasks vary over time.

A recent work introduces a stratified method for fog, edge, and cloud layers, with each layer employing a different technique [31]. Notably, much of the IoT-cloud research is silent on considerations like computational complexity and convergence rates, which become extremely important as processing moves into central clouds, particularly in the context of large-scale real-time applications. Addressing these concerns is important for collaborative edge-cloud systems.

Real-time industrial management and control requires the deployment of DNN models at multiple levels. In [32], researchers have described works that partition DNN over end devices, edge, and central clouds. They suggest that such a distribution is useful for inference performance when the computing resources of edge and end devices are limited and the remote transmission of data from these devices to clouds is costly. According to the authors, the optimization target for such a distributed system is described by the following relationships:

$$P_{total}^a = max/min\{\sum_{n=0}^{N}\sum_{}^{k_n}\{p_{n,i}^{r,a}\sigma(i,r) + p_{n,i}^{f,a}\sigma(i,f)$$
$$+ p_{n,i}^{l,a}\sigma(i,l)$$
$$+ p_{n,i}^{t,a}\prod_{j\in\{r,f,l\}}[\sigma(i,j)(1-\sigma(i+1,j))]\}\}$$

σ(i,r)+ σ(i,f)+ σ(i,t)=1 ∀i≤N

$$V_{b\neq ab=1}^{\alpha}\{\sum_{n=0}^{N}\sum_{}^{K_n}\{p_{n,i}^{r,b}\sigma(i,r) + p_{n,i}^{f,b}\sigma(i,f) + p_{n,i}^{l,b}\sigma(i,l)$$
$$+ p_{n,i}^{t,b}\prod_{j\in\{r,f,l\}}[\sigma(i,j)(1-\sigma(i+1,j))]\} \leq \hat{C}_b\}$$

Where P denotes the performance, subscript n, the nth type of partition, subscript/superscript a, the type-a performance, and $\hat{C}_b$ indicates the constraint boundary of performance b.

The type-n set of indicators, which include accuracy and delay time of inference, are defined by:

$$P^n = \{P_1^n, P_2^n, \ldots, P_\alpha^n\}.$$

To take the example of type-a transmission parameters we have:
$$p_n^{t,a} = \{p_{n,0}^{t,a}, p_{n,1}^{t,a}, \ldots, p_{n,k}^{t,a}\}.$$

Parameters for performance in the cloud and the edge are similarly defined and $P_a^n$ is the combination of all performance parameters. Then for all $\forall a \in \alpha$, type-a performance of type-n DNN is denoted as follows:

$$P_a^n = \bigcup_{i=1}^{k} p_{n,i}^{r,a} \vee p_{n,i}^{f,a} \vee p_{n,i}^{l,a} \vee p_{n,i}^{t,a}$$

where r, f and l denote cloud, edge, and end devices.

A representative function $\sigma(i, u)$ is is defined as follows to indicate whether layer i is on device u:

$$\sigma(i, u) = \begin{cases} 1, & if\ i \in u \\ 0, & if\ i \notin u \end{cases}$$

For more details of the model readers are referred to [31].

For a system with a central server and M local servers the researchers have implemented collaborative cloud-edge training based on federated learning [33]. Each edge server is trained on its local dataset. The set $\mathcal{E}$ of edge servers are mathematically expressed as
$\mathcal{E} = \{E_1, E_2, \ldots, E_M\}$.

The edge training datasets are:

$\mathcal{D} = \{D_1, D_2, \ldots, D_M\}$

Each edge server trains its local model **$m_i$** with dataset **$D_i$** using stochastic gradient descent (SGD). In round r+1 the edge server sets it local model **$m_i$** to the current global model **$m_g$**

$$m_{i,r+1}^0 = m_g^r$$

The local model training process at each step **s** can be expressed as:

$$m_{i,r+1}^s = m_{i,r+1}^{s-1} - \frac{\gamma_{r+1}^s}{N_{i,r+1}^s} \sum_{n \in D_{i,r+1}^s} \nabla f(m_{i,r+1}^{s-1}, X_{i,n}, Y_{i,n})$$

$\gamma$ is the learning rate and the size of the randomly selected dataset $D_{i,r+1}^s$ is $N_{i,r+1}^s$

After the $(r+1)^{th}$ round, the central server updates its global model by aggregating edge local models:

$$m_g^{r+1} = \frac{N_i}{\mathcal{N}} \sum_{i=1}^{M} m_{i,r+1}$$

Where $\mathcal{N}$ is the total number of samples aggregated over all edge models. The researchers claim marked improvement in dimensionality reduction and training efficiency over LSTM and GRU models. More details of experimental validation can be seen at [33].

*2.3 Security in Collaborative Edge-Cloud AI*

Systems integrated with the continuum face security issues that need to be carefully assessed. The presence of potential vulnerabilities arises from attackers exploiting IoT, edge and central cloud application programming interfaces to tamper with the parameters or inject malicious scripts. Such intrusions can result in improper model training and the loss of data. Additionally, devising robust security solutions for these systems is difficult due to the inherent heterogeneity of devices [34].

The authors in [35] confirm that critical systems are becoming increasingly dependent on cyber-based technologies, necessitating the use of models at multiple levels. The hierarchical attack detection process uses the results of the lower domain models' or extracted features to detect attacks in the upper domains, adding complexity to designing effective detection strategies. This requires the use of intra-domain and inter-domain communication networks. The communication technologies used in various domains include wired (e.g., LAN, WAN) [13], network devices (e.g., switches, routers), and wireless (e.g., Bluetooth, ZigBee, Wi-Fi, 4G, and 5G). The inter-domain and intra-domain communication networks are vulnerable to various kinds of attacks.

As described in the previous section, in many systems, the edge network takes the form of Multi-access edge computing (MEC). However, MEC is usually a distributed cloud with nodes physically located in many unsecured areas. The MEC platform may be housed at the base station or the radio network controller (RNC), thereby getting connected to the Internet. This significantly expands the attack surface of the system. Since the base stations are connected through RNC to the core network, hackers will have the opportunity to disrupt applications running on different network slices. Additionally, there are many other threats that pose challenges for edge-cloud-AI implementations, including vulnerabilities in communication over backhaul, manipulation of network software inventory records, hardware tampering, and attacks on APIs and AI models. These complications call for enhanced security of traffic on communication channels in various parts of the network.

Just as the training and inference workloads of the application are deployed over the continuum, the AI-based security workloads are also collaboratively handled over the continuum using the same principles of collaborative working. It is also necessary to acknowledge the complexity of handling security in CI, which involves ensuring the integrity of dataflows between IoT devices and the edge, as well as between the edge and the central clouds. This becomes an important aspect of overall security [36].

3. Challenges and Motivation

Following a detailed literature review, our analysis shows that researchers emphasize the superiority of collaborative, hierarchical IoT-edge-cloud-AI computing over using either edge or central clouds alone. This is due to the former's ability to distribute models, data, and task processing to provide superior outcomes [37], [38] resulting in better performance. Even though this paradigm is proving to be quite useful for critical infrastructure applications, it brings several challenges. In the following two subsections, we consolidate the issues that we have previously highlighted and propose strategies to address some of these challenges.

*3.1 Issues with the Current Approaches*

It is evident from the existing research that there are significant gaps in scientific knowledge relating to the security of critical infrastructure deployed over the IEC continuum, motivating our research efforts. The first level of complexity stems from the abundance and heterogeneity of markers, alerts, and measurement data generated by IoT, and other monitoring devices. This leads to the generation of datasets characterized by substantial volume and high dimensionality with intricate inter-feature relationships. The second level of complexity emerges from the complex nature of DNN models in the central cloud part of the IEC continuum. As discussed in Section 4, the polynomial time complexity of the fifth order of the computations associated with DNN models, exacerbates the challenges of training convergence. In real-time CI applications, the problem becomes even more severe as the edge and central cloud models require frequent retraining because of the constant flow of data from the operational technology domain. The third level of complexity is introduced by the integration of these systems with the Internet at multiple points, which expands the attack surface, escalating vulnerability to cyber threats. Researchers agree that dealing with model complexity necessitates the use of additional techniques for parallelizing processing and reducing complexity [39], [11], [40]. Attempts at reducing training time with these techniques often involve a trade off with detection accuracy [39]. Therefore, it is imperative to develop and evaluate innovative DNN model training techniques for the hierarchical and distributed IoT-edge-central cloud environment of CI systems.

This discussion will not be complete without mentioning Federated Learning (FL), which is a relatively new technique that trains deep learning models, like the DNN, in a distributed, collaborative manner without having to send local data to a central node. In the FL framework, the central cloud server initializes a model and distributes it to each edge node. The central model is then updated using parameters from the participating edge models. FL requires frequent exchange of model parameters between the cloud and edge, which becomes impractical for large-scale models. These exchanges are costly and hinder frequent transmissions, particularly with limited communication resources. Furthermore, model aggregation is designed for homogeneous models and does not account for the heterogeneity among edge nodes. The primary computational load of model training in FL is handled by the edge, placing a significant burden on these nodes. In contrast, the central server's task for aggregation remains relatively simple, leading to an imbalance in computational task distribution. However, due to limited resources, the aggregated model is difficult to deploy on devices.

In collaborative techniques like FL, effectively managing device and devices and data heterogeneity poses non-trivial challenge while handling security and privacy concerns [22], [26]. The multitude of sensors and other edge devices makes this approach more vulnerable to attacks than traditional deep learning methods [25], [41], [42]. An attacker may upload malicious data to the server and poison models, leading to a performance inferior to traditional algorithms.

*3.2 Significance of this Work*

Securing Critical Infrastructure (CI) is of utmost importance, and leveraging collaborative DNN models across the cloud continuum emerges as a promising approach to building robust security solutions. Existing methodologies present significant challenges. They either involve transmitting the central cloud model to edge clouds for localized training or developing models from scratch for both edge servers and the central cloud, both of which have several drawbacks.

Whether a model is downloaded from the central cloud to the edge or developed locally at the edge, attempts at running large models on edge clouds pose challenges. This situation necessitates the use of complexity reduction techniques like compression and pruning. However, these activities may be beyond the capabilities of the edge resources and result in an overload of resources, slowing normal training and inference processes and compromising accuracy. The execution of these tasks at the edge may also slow down its normal inference work and impact its accuracy. Additionally, iterative exchange of parameters between the edge and the central clouds results in high costs and latency, primarily due to constraints imposed by the communications network [43]. Methods involving many rounds of exchanging parameters between the edge and the central clouds increase communication costs [16], [27]. Moreover, despite data encryption, IoT-edge-cloud communication remains vulnerable to attacks.

With our innovative approach, we have achieved significant success in optimizing training time while preserving intrusion detection accuracy [44]. Our method introduces the Synthesized Learning System (SLS), a collaborative and hierarchical approach strategically designed to reduce the training time of large central cloud models. The SLS synthesizes central cloud models using trained layers from edge cloud models, addressing challenges present in conventional top-down distributed methods like FL [23], [24]. This avoids pushing large models from the central to edge clouds and consequently the challenges related to executing larger DNNs on edge resources.

One of the aspects that we have focused on is the effective mitigation of security risks associated with the integrity violation of data streams (including parameters of neural networks) flowing across the edge-cloud continuum. It would be naive to presume that exchange of gradients and other parameters in encrypted form can ensure system security.

While attackers may not be able to decipher encrypted data, they can still damage it, potentially leading to models being trained on corrupted data and wrong inferences with disastrous consequences. The SLS approach protects the dataflows against integrity violations from both known and unknown attacks. We also bridge the analysis gap in research by analyzing the computational complexity of the central cloud DNN models and their training convergence times.

These observations highlight significant research gaps that inhibit our ability to secure CI systems using the existing distributed multi-cloud AI techniques. It is important to address this research gap by effectively applying a distributed multi-cloud AI approach to enhance the security of CI systems. Through this paper, we seek to present a novel approach to using collaborative DNN models to protect the integrity of dataflows in CI systems that use IoT-edge-cloud architecture.

4. System Design

In this section we discuss the theoretical foundations and various design elements that form the basis of our work on the security of collaborative edge-cloud-AI systems. In Section 4.1, we provide an abstraction of the IOT-cloud based management control system within critical infrastructure systems. This abstraction reveals shared characteristics among CI systems, underscoring the relevance of the proposed security methods for security of dataflows across a wide range of CI systems. This model also leads to the security architecture that we describe in Section 4.3. The implementation of the networking in the IoT-edge-cloud continuum is discussed in Section 4.2.

*4.1 Conceptual Model of IOT Based CI Systems*

We have drawn from earlier works to abstract a conceptual model for managing and controlling IoT enabled CI systems. While the physical characteristics of CI system across different sectors may vary, their management control systems show similar functionalities. The metadata associated with dataflows, remains consistent across these diverse systems. In [45] the authors examine various scholarly works that conceptualize CI systems. In particular, the research in [46] highlights the importance of adopting an implementation-neutral conceptual model. Additionally, the authors in [47] introduce a comprehensive multi-tier model for critical infrastructure. A diagrammatic representation of the model is shown in Fig. 1.

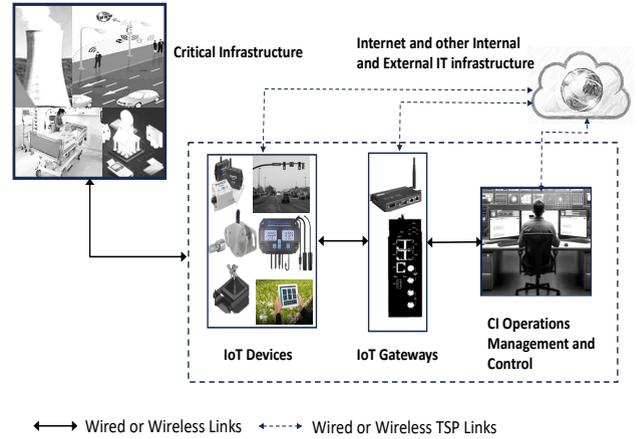

**Fig. 1.** Conceptual Model of IoT Based CI Systems

The IoT domain consists of strategically located sensor devices, both centrally and remotely, to monitor equipment and processes in the OT domain. Besides sensors, the IoT domain also contains actuators that might be triggered by the management control system (MCS, also known as supervisory or SCADA systems in industrial processes). The MCS acts based on the state of the system as indicated by the data collected in real-time. The altered state of the system is reflected in the metadata that is used to retrain the concerned edge models. IoT devices are connected to the gateways, serving as the crucial interface between the OT and IT. IT resources are largely provided by the edge and central clouds. These aspects are shared by most CIs. The MCS of the CI is an important part of the OT domain.

Another important component of this model is the network that links the devices in the IoT domain to the IoT gateways responsible for collecting and preprocessing. These gateways transmit the data to the MCS, which processes it further and uses it to control the operation of the CI system. The network may consist of wired and wireless links, owned or leased from telecommunication service providers (TSP). Since most CI systems generate an immense amount of high-dimensional data, the MCS relies on the resources provided by the IT domain for additional processing, including specialized processing that may need more powerful compute and storage resources than those available within the OT domain. This model clearly brings out the fact that since connectivity to the Internet extends to all tiers, attack surface increases and the risk of attackers attacking the inter-tier dataflows also increases.

*4.2 IOT-Edge-Cloud Tiered Architecture*

Integration of OT with IT through the IoT devices provides the much-needed processing, storage, and machine intelligence resources that the IoT devices lack. Fig. 2 shows the tiered model of the MCS of CI systems, separating out the OT and IT components into different domains. One prominent challenge that this architecture reveals is the vulnerability of the CI dataflows. Distinct trust boundaries exist between IoT and edge nodes (physical servers or clouds), and between the edge and central clouds. For proper functioning of CI systems, it is extremely important that the integrity dataflows traversing these boundaries should be protected. Additionally, given their connection to the Internet, clouds introduce vulnerabilities to cyberattacks that put the overall security of the system at risk.

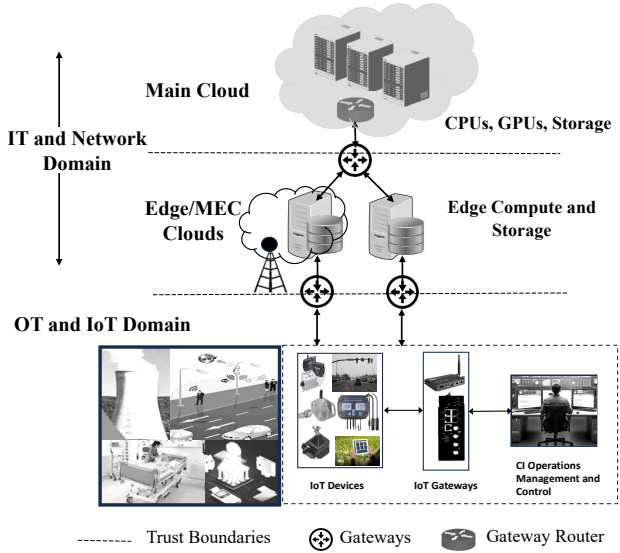

**Fig. 2.** CI MCS domains

The tiered architectural framework has allowed us to design a distributed security solution. Deploying AI models in both edge and central clouds allows for a collaborative approach to protecting dataflows, achieving a balance between how fast the security sub-system responds and its cost-effectiveness.

*4.3 Integration of MEC and 5G in the edge-cloud continuum*

The following standards form the basis for this discussion: ETSI GR MEC 031 V2.1.1, ETSI GS MEC 003: and the ETSO white paper on MEC in 5G networks [50-54]. These documents in turn make a reference to 3GPP 5G standards. In 5G, edge computing is identified as one of the key technologies required to support low latency together with mission critical and future IoT services. In integrating MEC with 5G, ETSI has focused on the architecture where the core network functions interact with each other using a Service Based Architecture [white], aligning system operations with the network virtualization and Software Defined Networking paradigms.

In 5G, the network functions and their services are registered in a Network Resource Function (NRF). In MEC the services are registered in the service registry. Network Exposure Function (NEF) is a 5G core cloud native network element that securely exposes the network services and capabilities, such as data and network services, to third-party applications over the Application Programming Interface (API) with assigned security and data integrity policies. Untrusted AFs that are not part of the operator's network, like that owned by cloud or edge service provider, are not allowed by the operator to directly access the target Network Functions. 3GPP 5G provides a common API framework for 3GPP northbound APIs (CAPIF). The relationship between MEC APIs and the CAPIF is shown in Fig. 3. The figure also shows how a clear path for data is provided through the UPF in 3GPP. The User Plane Function (UPF) has a key role in an integrated MEC deployment in a 5G network.

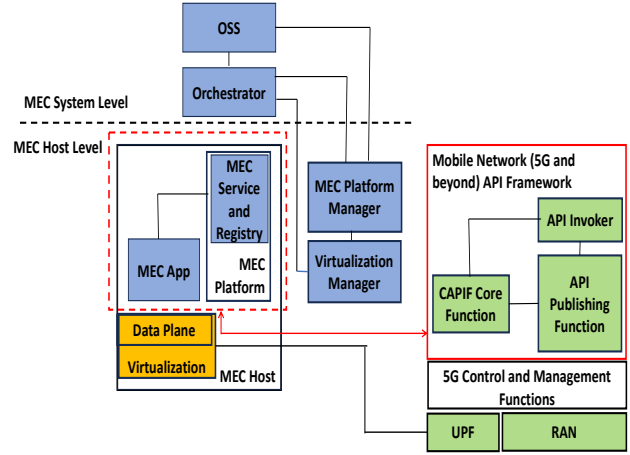

**Fig. 3.** Integrating MEC in 5G Network

In addition to AF, NEF and NRF, there are other 5G functions that are worth introducing. The procedures related to authentication are served by AUSF. Please see Table 1 for the full forms of these and other abbreviations. Network slicing is an important concept in 5G that allows One of the key concepts in 5G is Network Slicing that allows customized features and resources for different services or tenants using the services. Edge is part of the low latency slice covering 5G and MEC. This helps edge provide low latency services. NSSF is the function that assists in the selection of a suitable network slice instance and AMF for the users. A MEC application can belong to one or more network slices of the 5G network. The MEC orchestrator interacts with NEFs to obtain the required functionalities. MEC request the use of PCF to configure traffic steering rules for applications. Without going into the implementation details, it is worth mentioning that UPF plays an important role in the integration of MEC and 5G (see Fig. 3). The MEC host-level functional entities are usually deployed in a data network in the 5G system. Fig. 4 shows an example of the actual physical deployment of the MEC, in which MEC is collocated with the 5G base station.

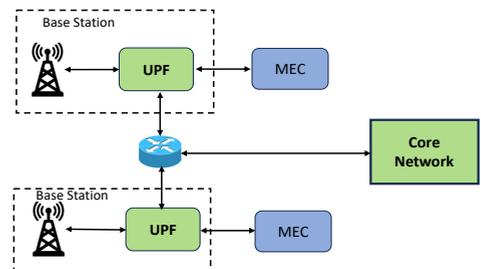

**Fig. 4.** MEC Collocated with the Base Station

5G networks can also integrate with cloud service providers to provide computing services or offload computations or storage. A possible configuration is shown in Fig. 5.

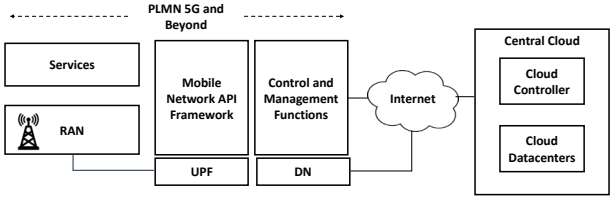

**Fig. 5.** Third party cloud integration with MEC in 5G

| Table 1 Acronyms Used | |
|---|---|
| Acronym | Expansion |
| AF | Application Function |
| API | Application Programming Interface |
| AMF | Access and Mobility Management Function |
| AUSF | Access and Mobility Management Function |
| CAPIF | Common API Framework for 3GPP northbound APIs |
| CN | Core Network |
| DN | Data Network |
| DNN | Deep Neural Network |
| CI | Critical Infrastructure |
| FL | Federated Learning |
| LDN | Local Data Network |
| MEC | Multi-access Edge Computing |
| MCS | Management and Control System |
| NEF | Network Exposure Function |
| NF | Network Function |
| NRF | Network Repository Function |
| NSSF | Network Slice Selection Function |
| OSS | Operation Support System |
| PCF | Policy Control Function |
| (R)AN | (Radio) Access Network |
| RBN | Restricted Boltzmann Machines |
| SCADA | Supervisory Control and Data Acquisition |
| SMF | Session Management Function |
| SLS | Synthesized Learning System |
| UCMF | UE radio Capability Management Function |
| UE | User Equipment |
| UDM | Unified Data Management |
| UDR | Unified Data Repository |
| UPF | User Plane Function |

This integration is made easier by the fact that IMT 2020 uses software defined network and network function virtualization. Interdependence of 5G and cloud computing has been growing and is bound to grow deeper. The PLMN constituents like mobile edge computing and fog are finding increasing roles in cloud computing. 5G in turn is undergoing cloudification through use of network function virtualization and software defined networks. This mutually interdependent and transformative relationship will drastically deepen with 6G and the future cloud computing with synergistic use of multi-cloud, quantum computing and communication and AI/ML integration. Beyond 2030 they are set to offer immersive, ubiquitous, and sensory digital experiences [55].

With the use of MEC in the edge, the 5G network provides a clear path to enable third party edge clouds like those owned by tower infrastructure companies and industrial facilities management. They interact with NEF to configure how application traffic in the user plane is directed towards MEC applications in data network. Finally, with the IoT, 5G and cloud integration the continuum looks as in Fig. 6.

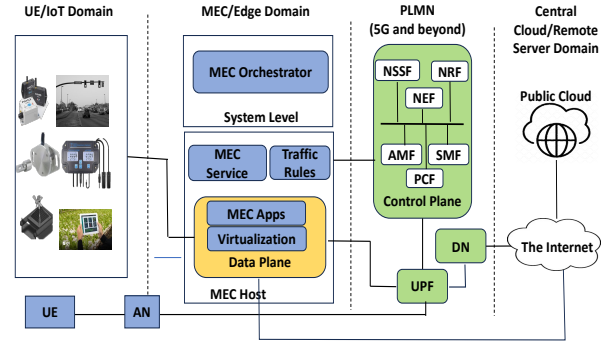

**Fig. 6** Complete integration of IEC continuum

### 4.4 The Security Architecture for CI Systems

Having seen the elements and complete integration of the IEC continuum, we are now ready to discuss the security architecture. This architecture spans the OT and the IT domain to provide security to the dataflows from the IoT gateways through the edge to the central cloud. A block representation of the architecture is shown in Fig. 7. It incorporates physical or virtual gateways at the trust boundaries separating various domains. These gateways perform a multifaceted role including data format and code conversion, policy implementation and participating in ensuring security.

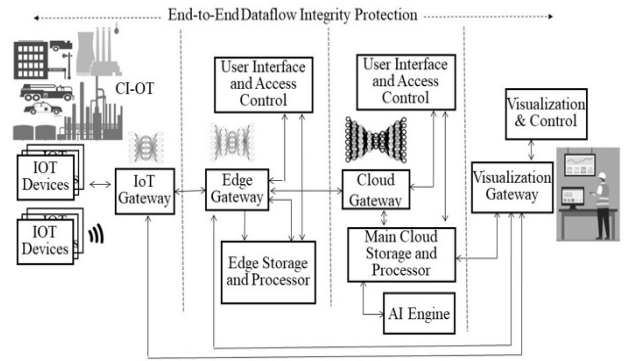

**Fig. 7.** Collaborative security architecture

Each gateway faces two attack surfaces. While the IoT gateway looks at the edge and IoT domains, the cloud gateway faces the edge and central cloud domains. Protection of models, dataflows, and tasks at these gateways is imperative to thwart potential attacks. The proposed architecture allows for development of our CI implementation agnostic security approach that leverages collaborative and hierarchical DNNs in a manner that aligns with the requirements of CI systems. It is noteworthy that the collaborative hierarchical architecture can be employed to process CI system operation workloads is also used for processing security workloads.

This architecture allows for creation, training, and usage of DNNs in the edge and cloud domains. By extension, simpler models can be trained and executed on IoT gateway. Metadata, such as source and destination byte count, IP addresses, or the source and destination and inter-packet times, extracted from the edge-cloud dataflows, are the features of the datasets used for training the edge-cloud models. In the proposed SLS methodology, each edge server/cloud model is trained with the data gathered through the IoT devices monitoring CI subsystems in its area. The central cloud model is synthesized using all or selected trained layers from the edge cloud models.

In the next subsection, we discuss the effect that training has on different layers of deep neural networks and how to evaluate layers for inclusion in the central cloud model. We show in the evaluation section that this layer selection method effectively reduces the training duration of the central cloud model. With such an arrangement, edge and central clouds can jointly handle a larger number of training instances and model parameters, thereby improving convergence rate and accuracy in intrusion detection. The experimental results, presented in the following sections, illustrate that this innovative arrangement optimizes the synergy between edge and central clouds, enhancing their combined capabilities for significantly improved intrusion detection performance. In one of the later sections, we compare the training performance of SLS with FL. Algorithms for edge cloud training, layer selection, and synthesis of central clouds are discussed in Section 5.

### 4.5 The Concepts of Layer Evaluation for Synthesis of Central Cloud Models

Because of the availability of abundant resources, the central cloud models are designed deeper and wider compared to the edge cloud models. This works well as complex jobs requiring large amounts of data are processed in the central cloud while low latency turnaround requirements can be met in the edge cloud. In distributed, hierarchical systems, the overall training and retraining times are dominated by the training times of the central cloud model. This presents a challenge when using distributed DNNs for CI systems that are generally complex yet time sensitive. Traditional deep learning models, including the federated models, may exacerbate this problem with their potentially high training times for complex central cloud models.

In the SLS approach presented in this paper, much like the traditional methods, the relatively smaller edge cloud models are trained with local data from their respective catchment areas. However, the difference lies in how the central cloud model is constructed and trained. In SLS, we use Algorithm 2 to take all or specific layers from the trained edge cloud models to synthesize a model for the central cloud. This results in a marked reduction in the training time of the central cloud model. This strategic methodological shift in the development and training of central cloud models addresses the challenge of long convergence times associated with complex central cloud models, offering a more efficient and responsive solution that meets the demands of CI systems.

We have observed through our experiments that there is a significant improvement in the training performance of central cloud models even if we do not use edge layers selectively but incorporate all trained layers from the edge models. Strategic selection of layers, based on their extent of contribution to the training of the model they belong to, further amplifies the advantage of the synthesized learning method. In [48] the authors analyze the layer-wise spectral bias of DNNs and relate it to the contributions of different layers in the reduction of generalization error for a given target function. In our study of the contribution of different layers towards the training of DNN, we have observed that different layers react to training differently. As the training process proceeds, some layers undergo large changes in parameters like weights, biases, and gradients and at the same time other layers hardly seem to change. The values of weights change during both the forward and backward propagation. Weights are indicative of the influence of the input data on the output and reduction of the generalization error.

Given a DNN with $n = N$ layers, and m neurons in each layer, we assume that $w_{ij}$ is the weight associated with the $j^{th}$ neuron in the $i^{th}$ layer.

$$\text{The sum of weights of all the layers} = \sum_i \sum_j w_{ij} \qquad (1)$$

$$\text{The sum of weights of the last hidden layer} = \sum_{i=(N-1)} \sum_j w_{ij} \qquad (2)$$

To get a sense of how the initial layers respond to training vis-a-vis the later layers, we compare the changes in weight in two ways:
i) By comparing the ratio of sum of weights of each layer to the total weight of all hidden layers (α_ratio).
ii) By evaluating the ratio of the sum of weights of each hidden layer to the sum of the weight of the last hidden layer (β_ratio).

These ratios can be expressed as follows:
For the $l^{th}$ layer, where $l = (1, n-2)$

$$\alpha\_ratio = \frac{\sum_j w_{lj}}{\sum_i \sum_j w_{ij}} \qquad (3)$$

$$\beta\_ratio = \frac{\sum_j w_{lj}}{\sum_{i=(N-1)} \sum_j w_{ij}} \qquad (4)$$

We will see the experimental results of how training affects different layers of a model in Section 5. The experimental results, presented in Section 5, illustrate that this innovative arrangement optimizes the synergy between edge and central clouds, enhancing their combined capabilities for significantly improved intrusion detection performance. In one of the later sections, we compare the training performance of SLS with FL.

*4.6 Complexity Analysis*

To keep the analysis simple, we assume that the neural network model has n layers, and each layer has n neurons. Each round (or epoch) of training has a forward and a backward propagation step. Following the analysis in [45], the number of multiplications in the forward propagation step is

$$n^{mult} = n^{layers} \cdot n^3 \qquad (5)$$

This gives the asymptotic run time of $O(n^4)$. Evaluation of activation functions is done element wise and has a time complexity of $T_a$, where:

$$T_a = O(n) \cdot O(n) = O(n^2) \qquad (6)$$

Factoring this in (5) we get the time complexity of the forward propagation $T_f$

$$T_f = O(n^4 + n^2) = O(n^4) \qquad (7)$$

The calculation of errors in the backward propagation has a time complexity of $T_b$, where:

$$T_b = O(time_{error}) = O(n^4). \qquad (8)$$

The calculation of weights has the complexity of $T_w$, where:

$$T_w = O(time_{weights}) = O(time_{error}) + n^3. \qquad (9)$$

If we factor in the calculation of n gradient descents, the overall complexity T, where:

$$T = O(n^5) \qquad (10)$$

We can see that the upper bound of the time complexity is a polynomial of order 5. In real life situation this would mean long training convergence times which multiplies as the value of n increases. We discuss in the next section on empirical validation how the method of central cloud model synthesis reduces the training time.

## 5. Empirical Validation

In this section, we describe our datasets, training and testing algorithms and the results showing acceleration in training time and preservation of accuracy of detection. We also compare the performance of the proposed technique with a collaborative non-synthesized model technique.

*5.1 Methodology*

1. Dataset Selection

After careful consideration, we opted for a diverse array of datasets encompassing network, IoT, and industrial IoT environment - specifically, UNSW (2015), BoT-IOT (2018), and TON-IoT (2021) [47], [46], [48]. These datasets contain instances of attacks capable of compromising the integrity of inter-domain dataflows. Notably, the TON-IoT dataset was gathered using IoT and IIoT sensors in the industry-4.0 environment.

To refine the performance of our models, we conducted feature evaluation and selection using Weka® 3.8.x to enhance model performance. We accounted for both similarities and differences between adjacent and distant edge cloud areas, subdividing the datasets accordingly. Each edge-cloud sub-dataset comprises 6000–10000 instances, with twenty percent reserved for testing. The models underwent multiple training epochs to consistently achieve an RMSE below the predetermined threshold, ensuring robust generalization.

2. Implementation Platform

We have tested the edge and central cloud models rigorously both on Mac and Windows machines. We have used, from time to time, Nvidia 1020, M4000M, and Nvidia GeForce RTX 2080B based GPUs. In the virtual environment on the Google Colab cloud platform, we have used other GPUs like the Nvidia K80 and P100, using the latest TensorFlow and Keras releases. The software environment consists of Keras with a TensorFlow 2 backend and Python 3.7 on the Anaconda platform. Some parts of the code have been ported to MATLAB.

3. DNN Hyperparameters and Training

We initialize weights randomly by drawing values from a uniform distribution within the range of -1 to 1, while setting all biases to 0. By initializing edge cloud DNNs with small random weights, the probability of the DNN falling into a trivial local minimum is greatly reduced. Subsequently, these parameters undergo changes during the neural network training process. In contrast, hyperparameters are meticulously selected to optimize performance. These included the depth and width of the DNNs, the choice of activation function and the learning rate. Stochastic gradient-based optimization was achieved using the Adam optimizer. The exponential decay rates ($ß1$ and $ß2$) were chosen for the Adam optimizer for best performance [49]. To determine the most effective values for these hyperparameters, an exhaustive search was conducted. The root mean square error (RMSE) serves as the loss function, providing a measure of how effectively the models learn from the datasets. Table 2 shows some of these hyperparameters and their typical values used in this work.

TABLE 2 DNN HYPERPARAMETERS

| Parameter | Description | Typical Value |
|---|---|---|
| Model depth | Number of layers | 4, 8, 12 |
| Model width | Number of neurons in each layer | 60, 120, 180 |
| Code Size | Innermost layer | 30 |
| Loss Function | RMSE | |
| $\lambda$ | Regularization factor | $10^{-8}$ |
| $\beta$ | Sparsity factor | 0.2 |
| $\beta1, \beta2$ | Decay rates | 0.8, 0.9 |

Table 3 describes the symbols used in the algorithms.

Table 3. SYMBOLS USED IN ALGORITHMS

| Symbol | Description |
|---|---|
| $X_{train}$, $X_{test}$ | Training and test data vectors |
| $X^k_{train}$, $X^k_{test}$ | Training and test datasets for edge cloud k |
| iter | Number of epochs of training |
| α_ratio | The ratio of sum of weights of each layer to the total weight of all hidden layers |
| β_ratio | The ratio of the sum of weights of each hidden layer to the sum of the weight of the last hidden layer |
| m | No of edge cloud models |
| $l_k$ | Number of layers in edge model k |
| iter | Number of training rounds for each edge cloud model |
| $S_{train}$ | Training dataset size of each edge model |
| $S_{test}$ | Test dataset size of each edge model |
| $W_{i,j}$ | Weight associated with the $j^{th}$ neuron of the $i^{th}$ layer |
| $S_{k,i}$ | Binary variable to denote whether layer i of model k has been selected |
| $t^C_k$ | Link between edge model k and central cloud model C |
| E | A set of trained edge-cloud models |
| $S_{k,i}.L_{k,i}$ | Includes layer i of cloud k if $S_{k,i}=1$ |
| $L_{k,i}$ | Layer i of the kth edge model |
| D | Synthesized central cloud model |

A trained DNN will give a low reconstruction error (RMSE) for normal instances while giving a high RMSE for compromised data streams.

---

**Algorithm 1.** Create, train and test edge-cloud DNNs

Define execution environment
Initialize dataset vector *X*
No of edge-cloud models ← *m*
No of layers in edge-cloud, $l_k$ ← *l*
No of training iterations ← *iter*
//split the dataset into train and test datasets
$X_{train}$, $X_{test}$ ← *split(X), ratio*(80 : 20)
//Prepare k non-overlapping sub-datasets for edge-clouds
$X_{train}$, $X_{test}$ ← *split(X), ratio*(80 : 20)
Initialize each train dataset size as $S_{train}$
Initialize each test dataset size as $S_{test}$
Initialize variables p ←1, q←$S_{train}$, r←1, s←$S_{test}$
**for** each edge-cloud $e_k$ k=(1,m) do //assign datasets
    $X^k_{train}$ ← $X_{train}$[p: q]
    $X^k_{test}$ ← $X_{test}$[r : s]
    p=q+1, q=(k+1)*$S_{train}$
    r=s+1, s=(k+1)*$S_{test}$
**end for**
**for** each $e_k$, k = 1,m **do** //train edge models
    Set hyperparameters
    Define input layer
    Define hidden layers
    Compile model
    Set epochs← *iter*
    **for** *epoch* = 1, *iter* do
        Train model using $X^k_{train}$ and validate model using $X^k_{test}$
    **end for**

---

**end for**

Training of smaller edge models usually proceeds fast taking around 20-50 epochs. These epochs are short in terms of time units. We will see the actual numbers later in this section. When the edge models have had a round of training, layers are selected for creation of the central cloud. Selection of trained layers is based on the discussion in subsection 4.4.

---

**Algorithm 2.** Selection of trained layers

Define execution environment
Use Algorithm 1 to obtain trained edge models $E_k$, k=1…M.
Depth of each model n ← N
Width of layers m ← M
Define binary variable S such that:
$S_{k,i} = \begin{cases} 1 & \text{if layer i of model k is selected for inclusion} \\ 0 & \text{if layer i of model k is not selected} \end{cases}$
Calculate α and β ratios for layers of models $E_k$
    **for** each model $E_k$, k=1, m **do**
        **for** each layer *l=1, m do*
            Initialize $S_{k,l}$ ← 0
            Calculate α and β ratios
            $S_{k,l}$ ← 1    \\if layer *l* included
            Determine $L_{k,l}.S_{k,l}$
        **end for**
**end for**

---

The synthesized central cloud model is already partially trained though the use of trained layers from the edge clouds. Training is completed using additional data from the central cloud area, particularly that which is not common with any of the edge clouds.

---

**Algorithm 3.** Algorithm for central cloud DNN model synthesis

Define execution environment
Define input shape
Initialize input layer weights
Call Algorithm 2
//Synthesize central cloud model from selected layers
$$D_{new} = \bigcup_{k=(1,m), i=(1,n)} S_{k,i} . L_{k,i}(X)$$

$D_{central} = input\_layer\ (\mathbf{x}.\mathbf{w}) || D_{new}$

Train and test model

---

5.2 Results And Discussions

We have carried out a comprehensive examination of the proposed synthesis method for training and the inherent layer selection technique. For training of the edge and central models, the datasets are prepared as discussed in the experimental setup in subsection 5.1. Each round of training, called an epoch, consists of one forward and one back propagation iteration of the complete dataset through the edge DNN.

1. Layer Evaluation and Selection for Model Synthesis

Drawing from our analysis in subsection 5, we test our hypothesis that different layers are affected differently during the training process. It thus makes sense to synthesize central cloud models by using trained layers of the edge models selectively. For both the cases, hidden layers of the same size and hidden layers of varying sizes, we observe in our study that the sum of the weights of all the hidden layers undergoes changes. We can see from Fig. 8(a) that the sum of weights shows a decreasing trend as the training progresses. The sum of weights of each of the four layers, as training progresses, is in Fig. 8(b). For the α_ratio described by Equation (3), we see from. Figs. 9 and 10 that layers 1 and 2 follow the training trend of the overall model. Layer 1 shows more reduction than layer 2 which indicates that layer 1 is affected more by the training process. For the β_ratio described in Equation (4), we see that as a proportion of the last hidden layer, weights of layer 1 undergoes the maximum changes followed by layer 2 (Fig. 6.)

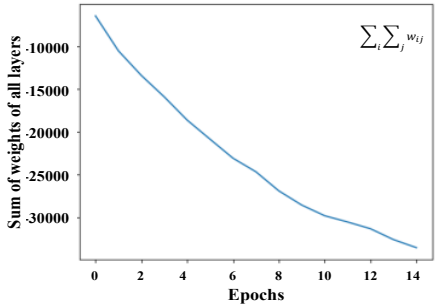

**Fig. 8 (a)** Sum of weights of all neurons (j) of all layers (i)

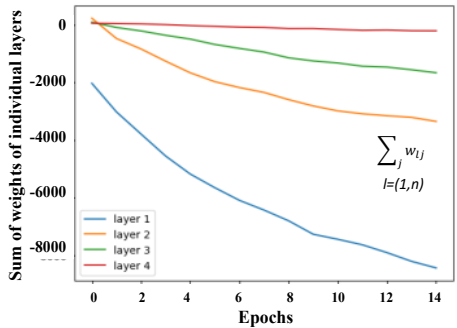

**Fig. 8 (b)** Sum of weights of individual layers

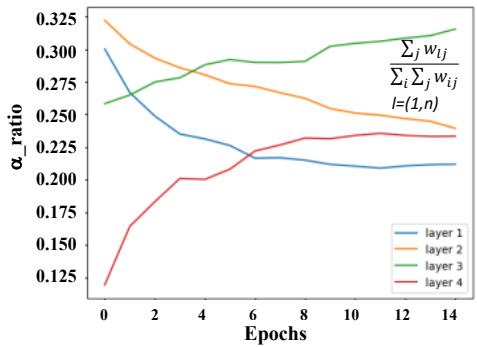

**Fig. 9.** Ratio of sum of weights of each layer and sum of the weights of the whole model

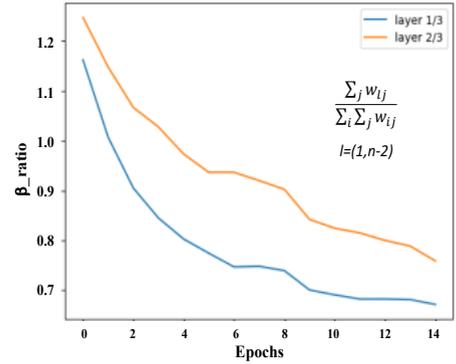

**Fig. 10.** Ratio of weights of each layer (*l*) that of the last hidden layer *l=(n-1)*

Based on our study, we conclude that the contribution of initial layers to the overall training process is comparatively much higher than that of the later layers. It is also observed that this process preserves the detection accuracy of integrity-violating attacks on inter-domain dataflows. As can be seen, this is different from FL in which identical models are trained in the central and edge clouds. It is also different from transfer learning as well, which involves a model trained for one situation being used for a different problem. We will see in a later subsection how this technique is used to construct synthesized central cloud models that train faster than models that do not use layers from the edge clouds.

2. Training and testing of edge and central cloud DNNs

We tested the edge, central and synthesized central clouds in two set-ups, A and B (Table 4). In Set up A we had edge models of depth 4 and with varying from 30 to 90 neurons. The idea was to test whether the synthesized central cloud converge significantly faster than the on that are constructed ab-initio.

| TABLE 4 EXPERIMENTAL SET-UPS FOR TESTING CONVERGENCE | | |
|---|---|---|
| **Set-up** | **Edge** | **Central** |
| Set-up A | {90, 60, 30, 30, 60, 90} | {180,90,60,30, 30 60,90,180} |
| Set-up B | {7*60}} | {10*60} |

Set-up A

The edge cloud models were trained on their respective datasets according to Algorithm 1 in subsection 5.1 clause 3. As illustrated in Figure 11, the losses decrease as training progresses. After conducting several runs, we observe that the minimum training and test losses for edge cloud 1 are 4.7, and 17.6, respectively, with corresponding accuracies of 0.993 and 0.995. In the case of edge cloud 2, the minimum losses are 8.5 for the training dataset and 8.7 for the test dataset, with corresponding accuracies of 0.98 in both cases. The losses of edge cloud models on the test data settle down to low values, close to the training losses, within 40-60 epochs of training. This convergence to low test losses is indicative of robust generalization, proving the models' ability to perform well on previously unseen data.

Despite its increased depth and width, the synthesized central cloud model exhibits a faster training pace compared to the considerably smaller edge cloud models in terms of the number of epochs required for training. In our experiments, training of the synthesized central model required only 3 to 10. This implies a remarkable acceleration in the training process of the central cloud when it is synthesized using trained layers from the edge clouds. In contrast, without reusing the trained edge cloud layers, the central models typically require 40-60 epochs to train, which translates into hours or days of training

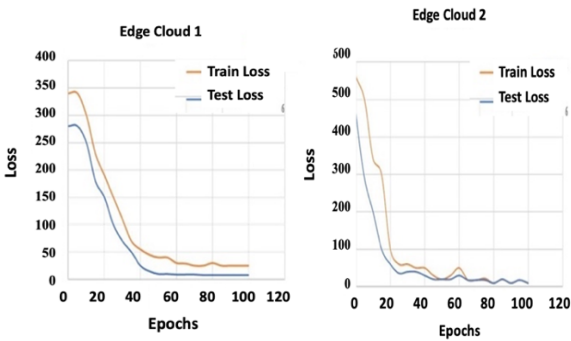

**Fig. 11** Training and testing of edge cloud models- set up A depending on the size of the model and that of the dataset. Fig. 12 illustrates representative training and testing results for the central cloud when all the layers of the edge clouds are used in the central cloud. Both train and test losses converge close to 5. The train-accuracy and test accuracies are close at 0.992 and 0.993 respectively. These outcomes underscore the efficiency gained by integrating pre-trained edge cloud layers into the central cloud model, thereby enhancing the overall performance.

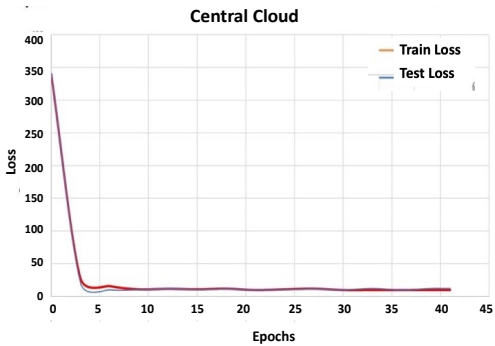

**Fig. 12** Training and testing of central cloud model – set up A

Set-up B

In set up B we altered the platform and using a subset of data prepared earlier to repeat the edge cloud training. This served the purpose of a more in-depth study of the edge cloud training and to prepare models of appropriate size from which layers could be selected to prepare central cloud models. As expected, the general the general training pattern is same as in Set-up A. We can see in these examples that training converged around 20 epochs. For illustrative purposes, for two of these curves we have enlarged the sections in which the convergence point lies.

Finally, we tested the layer selection method for constructing central cloud models to confirm that they converge faster than the non-synthesized central cloud models. Central cloud models had a depth of 10 layers while the edge cloud models were 5 layers deep. Various width sizes were experimented with for optimality. Results for training of some of the edge models are given in Fig 13 with the areas of convergence highlighted.

For comparison purposes, central cloud models were created ab-initio without using trained layers from the edge clouds. Convergence is usually in the same epoch region as the edge clouds. In set B this is between 20-30 epochs (Fig. 14)

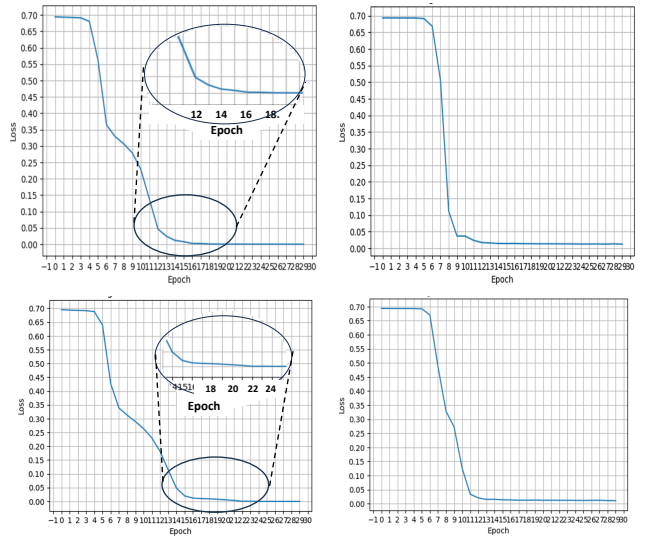

**Fig. 13** Training of the edge clouds – set up B

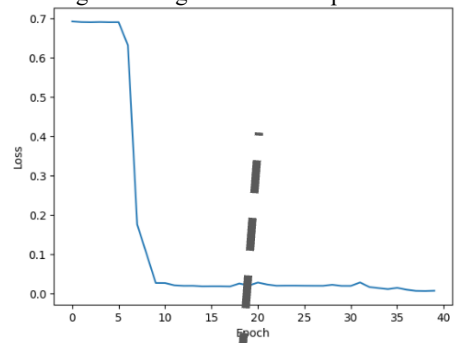

**Fig. 14** Training of the central cloud – Set-up B

The revelation begins when we start evaluating edge cloud layers for training efficiency and choosing different combinations of edge cloud layers to form central cloud models. True to our analysis in subsection 5.4, different combinations produced models with different convergence times. We have shown the worst and the best cases from our experiments in Figs. 15a and 15b. Worst case (longest convergence time) was for the central cloud with layers 1, 3, 4, from edge cloud 1 and layers 9, 11, 12 from edge cloud 2.

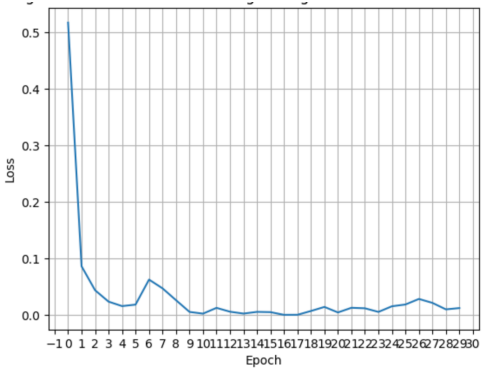

**Fig. 15a** Synthesized central cloud model (Worst case)

Fig. 15 shows that the best outcome (central model with fastest convergence) was seen with layers 1, 2, 3 and 5 from edge cloud 1 and layers 10 and 12 from edge cloud 2. In the best-case convergence happens close to epoch 3.

This was not the only combination that converge at 3. However, the notable point is that while a fresh central cloud model converges after about 20 epochs, the synthesized model converges after 3 epochs. What this means in terms of time, we will see in clause 5 of this subsection.

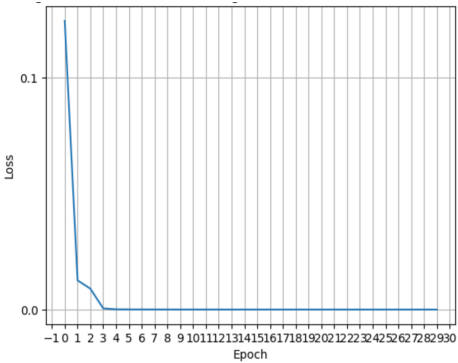

**Fig. 15b** Synthesized central cloud (Best case)

3.  Analysis of the Training time

In Section 4, we demonstrated that DNNs exhibit a time complexity of the order of $O(n^5)$. As models grow in terms depth and width and the number of training instances increase, the theoretical training time becomes impractically high in CI systems with complex central cloud models and large datasets. One of the few papers that discusses training times for such models is referenced in [41]. In their work, Santara et al. conducted baseline experiments with a 5-layer model having the largest dimension set at 1000. Their method involves 20 epochs of pre-training the Restricted Boltzmann Machine (RBM) followed by 10 epochs of fine-tuning by back propagation. Layer wise training for 20 epochs of forward and backward training takes 5 hours, 31 minutes, and 42 seconds. With synchronized pre-training, the total time is reduced to 4 hours, 4 minutes, and 53 seconds.

In our context, the edge cloud models, with a depth of four, require approximately 125 seconds (between 35 and 40 epochs) for 6000 training and 1200 test examples. Using our proposed method of synthesizing central clouds reduces the number of training epochs to around 6. In terms of time efficiency, this approach takes significantly less time than for a fresh stand-alone model. We compare training time of a baseline 8-layer model and a synthesized 12-layer model with parameters detailed in Table 5.

**TABLE 5** COMPARISON OF BASELINE AND SYNTHESIZED MODELS

|  | 8 Layer traditional model | 12-layer synthesized model |
|---|---|---|
| Layers | 180,90,60,30, 30 60,90,180 | 60x3,60x3,60x3,90x2 |
| Trainable parameters | 31521 | 27275 |
| Time for 100 epochs | 1 hour 24 minutes and 30 sec | 1 hour and 2 minutes |

The training times are reported for 7000 training and 1500 test examples. After multiple runs, it is evident that the synthesized 12-layer model takes 23.9% to 34.69% less time than even a considerably smaller 8-layer unmerged model.

For set-up B, we used edge models with 7 hidden layers and central cloud models with 10 hidden layers of various widths. We show in Table 6 results for edge models with 60 neurons each and central cloud models with 360 neurons each.

**TABLE 6.** COMPARISON OF CONVERGENCE TIME OF SYNTHESIZED WITH NEW CENTRAL CLOUD MODEL

|  | Edge | Central Cloud Model | Synthesized Central Cloud Model |
|---|---|---|---|
| Configuration | {7*60} | {10*360} | {10*360} |
| Time 40 epochs (4000 instances) | 15.26 s | 144 s | 84.2 |
| Epochs to converge | 11 | 14 | 11 |
| Time 40 epochs (20000 instances) | 63.62 s | 256.14 s | 222.04 s |
| Epochs to converge | 7 | 7.5 | 5 |
| Time to converge |  | 44.38 s | 31.23 s |

For comparison purposes to keep the training time within practical limits, training and testing of central cloud model was first done with 4400 and 1000 instances and for edge models with 1000 training and 300 test instances. A larger dataset consisting of 20,000 instances was then employed to train the models to analyze how these models will fare in real-life systems. For a width of 60 neurons and dataset with 1000 training instances, the edge model takes 15.26 seconds for 40 epochs. It, however, converges in 11 epochs. Comparatively synthesized central cloud model with 60 neurons in each of the ten layers takes around 19.8 seconds for 40 epochs but converges in 7 epochs. This is because the synthesized model uses trained layers from the edge models. It is also observed that a new central cloud model with a width of 360 neurons takes, on an average, 253.60 seconds to complete 40 epochs. It converges in 7.5 epochs, averaged over many runs. On the other hand, a synthesized central cloud model of the same size takes about 249.15 seconds but converges in 3-6 epochs for different combination of layers. Statistical average of several runs yields a value of 5 epochs. The improvement in terms of epochs and convergence time are 29.62% and 33.33%. These results are shown in Table 6.

4. Performance of the synthesized model in detecting intrusions

To evaluate the intrusion detection capabilities of our proposed system, we utilized datasets comprising both normal and anomalous instances to assess the intrusion detection performance of our proposed system. We purposefully selected intrusions that alter the metadata and potentially compromise the integrity of the dataflows. We executed the trained models on the datasets described, drawing meaningful inferences.

Instances from dataflows affected by attacks yielded results with notably high RMSE while the normal instances were reconstructed with consistently low RMSE. Through a thorough experimental analysis, we establish a critical threshold value for effectively distinguishing between normal and affected data. Table 7 presents various results averaged over multiple runs using this threshold.

TABLE 7. ATTACK DETECTION PERFORMANCE

| Sl. No | Attack as Attack (TP) | Attack as Normal (FN) | Normal as Attack (FP) | Normal as Normal (TN) | Total Vectors | Accuracy (%) |
|---|---|---|---|---|---|---|
| 1. | 200 | 0 | 19 | 355 | 574 | 96.69 |
| 2. | 188 | 0 | 0 | 511 | 699 | 100.00 |
| 3. | 96 | 0 | 3 | 232 | 331 | 99.09 |
| 4. | 92 | 16 | 0 | 184 | 292 | 94.52 |
| 5. | 100 | 0 | 0 | 280 | 380 | 100.00 |
| 6. | 80 | 17 | 0 | 243 | 340 | 95.00 |
| 7. | 110 | 1 | 1 | 287 | 399 | 99.50 |
| 8. | 27 | 1 | 1 | 100 | 129 | 98.45 |
| 9. | 126 | 12 | 0 | 398 | 537 | 97.76 |

It is important to have high precision since normal traffic misclassified as an attack triggers potentially expensive investigation. Our model achieves a high precision of 96.5%, attesting the model's proficiency in identifying actual attack incidences. In any intrusion detection system, high false negative rate could allow attacks to go unnoticed leading to disastrous consequences. However, in our case the true positive rate is high at 98.6%. The false positive rate is 0.64%, indicating a low probability of false alarms.

5. Performance comparison with traditional methods

To put the performance of our method in proper perspective, we carried out a comparative study with FL, a contemporary approach that uses distributed architecture of deep neural networks and employs collaborative learning. In FL framework, the central model is created at the central cloud and subsequently transmitted to the edge clouds. Models are trained at the edge clouds on local datasets and parameters are exchanged with the central cloud.

Our FL set up consists of four edge or client models and one central model. To thoroughly evaluate performance, we configured models with varying layer configurations ranging from 3 layers to 12 layers as outlined in Table 8.

TABLE 8. CONFIGURATIONS OF FEDERATED LEARNING MODELS

| Layers | Configuration |
|---|---|
| 3 | {180,180,180} |
| 8 | {180, 90, 60, 30, 30, 60, 90, 180} |
| 12 | {180, ….180} |

Several iterations of the experiment were conducted with each iteration consisting of 40-50 rounds of training of the combined edge and central cloud setup. Each round consisted of 20 epochs of training for the edge cloud and exchange of parameters with the central cloud. Each of the experiments are repeated twenty times ensuring robust evaluation. Edge and central cloud learning rates have been chosen as 0.00000001 and 0.005. Kernel and bias use l2 regularization with regularization parameter at 0.0001. These parameters were selected by grid search for optimizing the outcome.

In Figs. 16 and Fig. 17 we show the loss performance of the models with different number of layers. We can see from the figure that losses flatten out between 40 and 50 epochs. The accuracy achieved is about 83.6%.

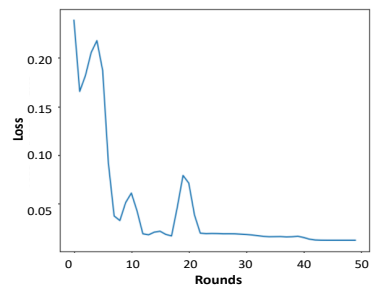

**Fig. 16.** Training of 3-Layer FL model

To compare the performance of our other traditional and synthesized models described in Table 7, we set up FL with 8 and 12-layer models. The number of clients remain four. Training performance is shown in Fig. 17 a) for the 8-layer model and Fig. 17 b) for the 12-layer model.

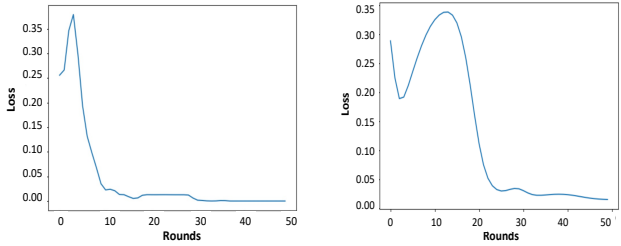

**Fig. 17a)** 8-layer FL models     Fig. **17b)** 12-layer FL models

Losses settle down around between 40 to 50 rounds of training with 8-layers model giving an accuracy of 98.3% and 12-layers model 97.5%. The corresponding training times are approximately 34 minutes, 39 minutes and 1 hour and 13 minutes. These are indicated in Table 9 below:

| TABLE 9 TRAINING TIMES OF FL MODELS ||
|---|---|
| Layers | Training Time |
| 8 | 2626 seconds |
| 12 | 4378 seconds |

It is seen that FL performs better than a system of models where the central cloud model is trained without any assistance form the edge cloud models. The time taken by FL comes to 43 minutes and 46 seconds as against 1 hour and 24 minutes of the non-collaborative models. It has thus performed better in this case. In contrast, when we compared the 12-layer SLS and FL systems, it is seen that FL takes 1 hour 13 minutes as against 1 hour and 2 minutes in the synthesized model. As compared to our synthesized model, the 1 hour 13 minutes taken by FL is about 18% higher. For larger models and datasets, this difference becomes important. We also observe that the training times in FL are quite sensitive to the number of neurons in different layers, more than the training times in the synthesized model. Thus, in real-life CI systems, as the models grow in width and depth, FL is more adversely affected than SLS in terms of convergence time. Our calculations do not account for the communication delays, which are generally higher in FL as evidenced earlier in the significance section of this paper.

6. Discussions and Conclusions

In conclusion, the complex nature of CI systems requires sophisticated management control mechanisms to ensure seamless operation. The quest for increasing automation is driving the integration of IOT with operational technology (OT) of the CI systems. This integration of heterogenous IoT devices results in generation of a vast amount of heterogenous and high-dimensional data. While the resulting IoT-edge-cloud continuum allows for innovative approaches to handle CI workloads, it concurrently exposes dataflows to expanded attack surfaces. Without proper mitigation, these attacks pose a risk to the integrity of data flows, potentially leading to failure of delivery of essential services. Experts believe that deep learning outperforms any other method for identifying sophisticated attack patterns. Unfortunately, currently available threat and attack detection approaches that use deep neural networks turn out to be unsuitable for real-time operations A critical consideration is the extended training times of DNN models in central clouds, which, if not addressed, could limit their applicability in CI systems that require real- or near-real-time responses. This paper addresses these challenges and introduces a novel solution called the synthesized learning system (SLS), a collaboratively trained system of DNN models in the central cloud the edge nodes to enhance the security of inter-domain dataflows. Our proposed approach involves synthesizing the central cloud model collaboratively by incorporating trained layers from edge cloud models, thereby significantly accelerating the training process, and improving convergence rates. This innovative solution holds promise for fortifying the security of critical systems. Our experiments demonstrate a remarkable improvement of up to 40% improvement in the training process over non-collaborative models and similar improvements over FL models. At the same time SLS preserves or enhances intrusion detection accuracy, with values ranging from 96.9% to 99.5%. Additional metrics, including precision and false positives, support and complement these findings.

We appreciate the importance of speed as well as accuracy of security implantations in CI systems. In this regard, we diversified our work in two critical directions - transparency in decision making by AI model and hybrid quantum edge computing. Transparency will empower CI operations and management personnel to appropriately query AI model decisions, not only improving confidence in AI based decision making but also offering valuable insights for continual model enhancement. Hybrid quantum edge promises to integrate classical and quantum resources to improve processing and communication latencies to the extent not possible by classical edge alone.

ACKNOWLEDGEMENT

This research has been supported in part by the Broadband and Cybersecurity Funding from the UM System Extension Program.